\documentclass{article}
\usepackage{epsfig}
\usepackage{latexsym}
\usepackage{amssymb}
\begin{document}

\title{Role of surface states in the Casimir force between semiconducting films}
\author{M. Govoni$^{1,*}$ \and C. Calandra$^{1}$ \and A. Benassi$^{2}$}

\maketitle
\begin{description}
\item[]  $^1$Dipartimento di Fisica, Universit\`a di Modena e Reggio Emilia
Via Campi 213/A, I--41125 Modena, Italy
\item[]  $^{2}$CNR/INFM--Natl. Research Center on nanoStructures and bioSystems at Surfaces S3
Via Campi 213/A, I--41125 Modena, Italy
and
CNR/INFM--Natl. Simulation Center Democritos
Via Beirut 2--4, I--34151 Trieste, Italy
\item[]  $^*$E-mail: marco.govoni@unimore.it
\end{description}
\begin{abstract}
We present results of first principle calculations of the Casimir force between Si films of nanometric size, which show that it depends significantly upon the configuration of the surface atoms, and give evidence of the importance of surface states.
\end{abstract}

\section{Introduction}
Recently theoretical results have been reported which show the influence of the thickness on the Casimir force between thin slabs\cite{Pirozhenko08,Lenac08,Benassi09}. These studies are motivated by the role that thin films play in micro- and nano-devices. They rely on two approximations: (i) a local description of the dielectric properties of the material, (ii) the use of a plasma or Drude--Lorentz model to represent the bulk dielectric function. In Ref.~\cite{Pirozhenko08,Lenac08} this same dielectric function has been adopted to represent the film dielectric properties.\\  
In this paper we derive the dielectric tensor of a Si film starting from the microscopic description based on Density Functional Theory (DFT) in a pseudo-potential approach, thus automatically including size quantization and surface state effects. Here we present preliminary results on the forces between Si films of nanometric size.

\section{Simulation of the Silicon film}
For intrinsic Si the optical excitations at low energy involve the valence electrons. We are interested in determining how they are modified in a film of small thickness compared to the bulk due to: (i) the confinement of the electron gas (Quantum Size Effects, QSE), (ii) the presence of surface states (Surface Effects, SE).\\
The system we investigate is a film obtained by stacking 12 atomic planes of Si along the $[111]$ direction. We take the surface normal as the $z$-axis. The film has 2D periodicity along the planes. If we terminate the film without allowing any rearrangement of the surface atoms the film turns out to be 1.9 nm thick and the surfaces show an half filled dangling bond per unit cell. On the other hand if we allow structural rearrangements at the surface, $2\times 1$ surfaces may be generated, with two dangling bonds per unit cell. Indeed the Si$(111)$ surface is the cleavage surface and, after cleavage, it shows a $2\times 1$ reconstruction\cite{Pandey81,Pandey82}, which produces zig-zag chains of dimers along a planar direction, that we take as the $y$-axis. In this case the film has a different two-dimensional periodicity and a smaller thickness (1.7 nm).
\begin{figure}[h]
\begin{center}
\epsfig{file=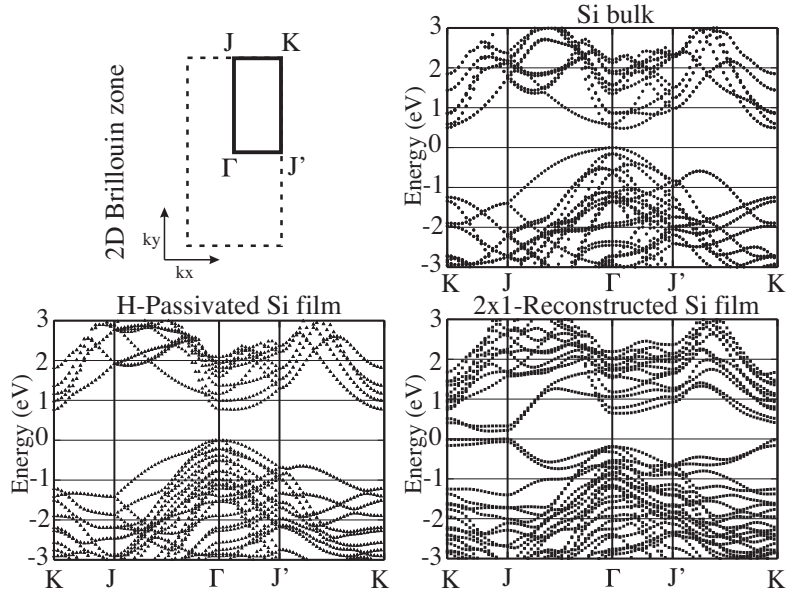,width=0.85\textwidth}
\end{center}
\caption{Energy band structure of  the H--passivated Si(111)  film and the one with $2\times 1$ reconstructed surfaces along the high symmetry directions of the 2D Brillouin zone. A sketch of the projected bulk band structure is given for comparison.}
\label{mg:fig1}
\end{figure}
One way to consider QSE only is to saturate the dangling bonds by atomic H adsorption, which removes the surface reconstruction and leaves the band gap free of surface states. Therefore one can separate the SE from QSE just by comparing the electronic structure of the H terminated film with the one of the film with the $2\times 1$ surfaces.\\
To calculate the film electronic structure we use the so-called repeated slab approach\cite{Schlueter78} and the DFT--KS implementation of the PWscf code\cite{QE-2009} in the Local Density Approximation (LDA).\\
\ref{mg:fig1} displays the 2D band structure along the high symmetry directions of the 2D Brillouin zone of the film with $2\times 1$ surfaces. We give also the band structure of the film with the H--passivated surface and a sketch of the bulk band structure. One can see that while the passivated surface shows a behaviour in the band gap region rather similar the bulk, the band structure for the film with $2\times 1$ surfaces has four bands in the band gap. The bands arise from surface states at the two sides of the film: one band of each couple is empty and the other is filled. These bands are responsible of the optical transitions that have been detected several years ago by surface optical absorption\cite{Pandey82}.

\begin{figure}[h]
\begin{center}
\epsfig{file=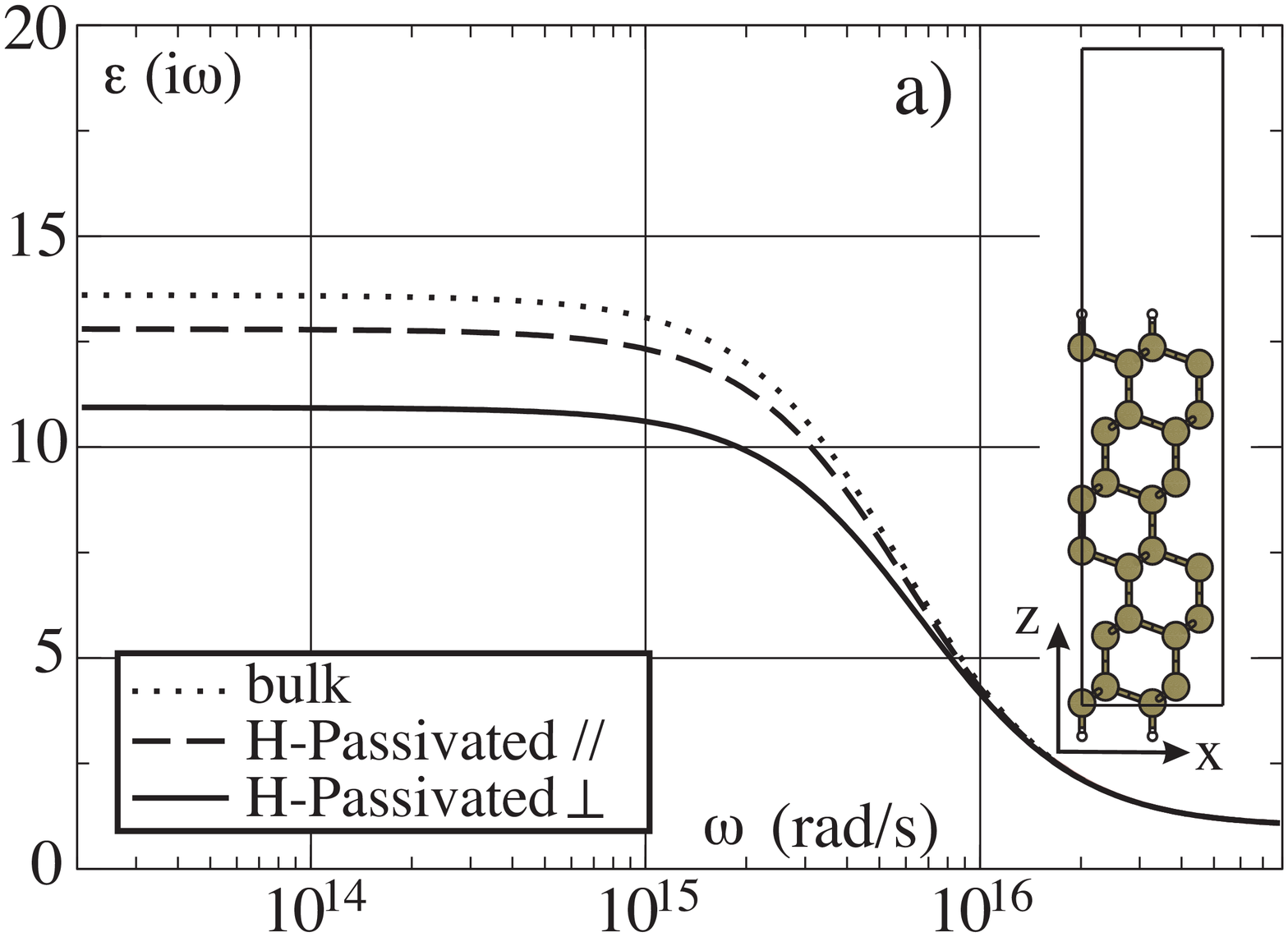,width=0.49\textwidth}
\epsfig{file=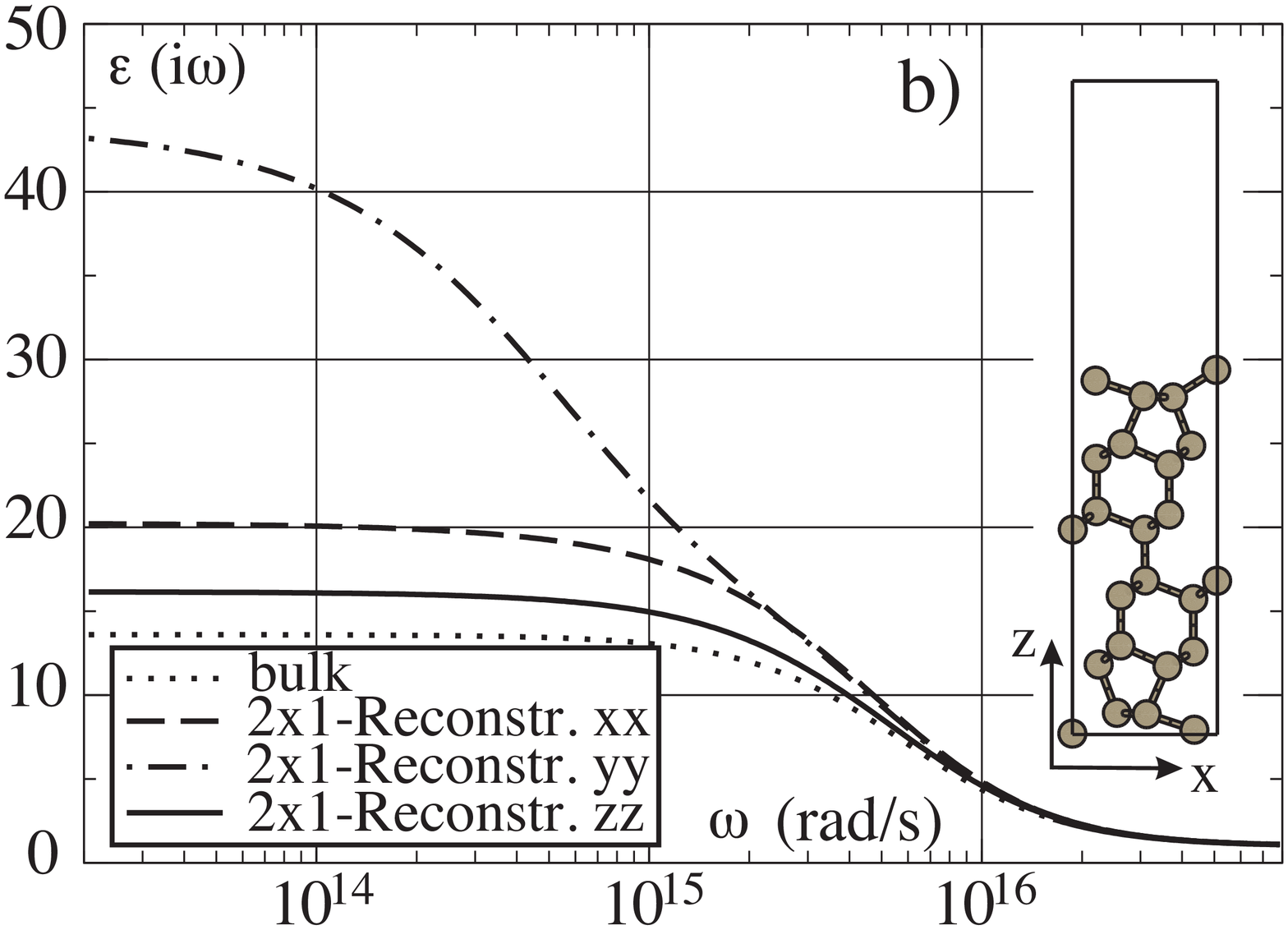,width=0.49\textwidth}
\end{center}
\caption{London transform of the components of the calculated macroscopic dielectric tensor. 2a) case of H--terminated film; 2b) film with $2\times 1$ surfaces. The insets show the film geometry.}
\label{mg:fig2}
\end{figure}
\section{The film dielectric tensor}
Having at disposal the electron energies and wavefunctions we have calculated the microscopic dielectric matrix\cite{SAX} and from it we have determined the macroscopic dielectric tensor of the film following the procedure indicated in Ref~\cite{Gavrilenko96}. In the case of bulk Si the dielectric tensor is isotropic and the diagonal comp1onents are given by the Si dielectric function. For the H--passivated film the tensor character arises essentially from size effects: the modifications induced by the H--Si bond in the electronic structure do not affect significantly the dielectric behaviour in the low energy range compared to the bulk. As a consequence of the symmetry the lateral components are equal $\epsilon_{xx}=\epsilon_{yy}=\epsilon_{\parallel}$ and take values close to the bulk dielectric function, while $\epsilon_{zz}=\epsilon_{\perp}$ is different. The London Transform of the dielectric tensor components are displayed in \ref{mg:fig2}a. It is seen that the values are slightly lower that the bulk dielectric function.\\
The situation changes drastically for the $2\times 1$ reconstructed film. As shown in \ref{mg:fig2}b the components of the dielectric tensor are higher than the bulk, the discrepancy being much larger for $\epsilon_{yy}$. The deviations with respect to the bulk are due to the presence of the surface bands, which reduce the band gap value leading to an increase of the static limit. 

\begin{figure}[h]
\begin{center}
\epsfig{file=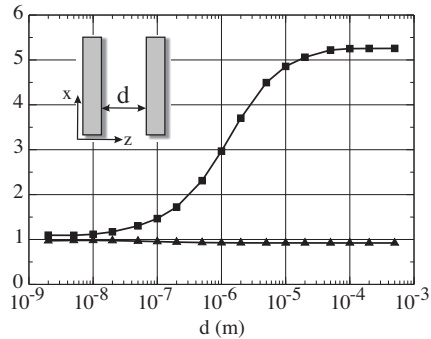,width=2.2in}
\end{center}
\caption{Plot of the ratios of the calculated Casimir forces between Si films as a function of the film distance. The curve with $\blacktriangle$ (\tiny$\blacksquare$\footnotesize) gives the ratio between the force calculated for the H--passivated ($2\times 1$ reconstructed) film and the one obtained with the bulk dielectric function.}
\label{mg:fig3}
\end{figure}
\section{Calculation of the Casimir force}
The calculations of the Casimir force can be done in the framework of the Lifshitz theory. For the H--passivated films this requires the extension of the formalism of Ref.~\cite{Zhou95} to films of uniaxial dielectric media. For the reconstructed surface one has to resort to a more complex theory\cite{Yeh88}. A detailed discussion of the extension of Lifshitz theory to this case will be given elsewhere\cite{Govoni10}.\\
To illustrate the importance of QSE and SE we give in \ref{mg:fig3} the calculated Casimir force per unit area between H--passivated and reconstructed Si films as a function of the distance. We plot the ratio between the force determined using the appropriate dielectric tensor and the one obtained with the bulk dielectric function. The comparison shows that the effect of surface states is quite significant for distances larger than 10 nm and becomes very important at large distances, where the force is essentially determined by the value of the London transform at low frequencies. We believe that these results give evidence of the necessity of accounting for the modifications of the electronic structure in the determination of the force between nanometric objects. 

\section*{Ackonwledgments}
M. Govoni acknowledges financial support by the European Science Foundation (ESF) within the activity `New Trends and Applications of the Casimir 
Effect' (www.casimir-network.com).

\bibliographystyle{ws-procs9x6}
\bibliography{QFEXT09_Marco_Govoni}

\end{document}